\documentclass[aps,showpacs,preprint,amsmath]{revtex4}
\def\bea{\begin{eqnarray}}
\def\eea{\end{eqnarray}}
\def\be{\begin{equation}}
\def\ee{\end{equation}}
\usepackage{graphicx}

\begin{document}

\title{Optimal reflection-free complex absorbing potentials for quantum propagation of wave-packets}

\author{Oded Shemer, Daria Brisker and Nimrod Moiseyev\footnote
{correspondence  author, nimrod@tx.technion.ac.il}}

\affiliation{Department of Chemistry and Minerva Center of
Nonlinear Physics in Complex Systems Technion -- Israel Institute
of Technology Haifa 32000, Israel. \\}

\begin{abstract}
The conditions for optimal reflection-free complex-absorbing
potentials (CAPs) are discussed. It is shown that the CAPs  as
derived from the smooth-exterior-scaling transformation of the
Hamiltonian, [J. Phys. B. {\bf 31},  1431 (1998)], serve as
optimal reflection-free CAPs (RF-CAPs) in wave-packet propagation
calculations of open systems. The initial wave packet, $\Phi(t=0)$
can be located in the interaction region (as in half collision
experiments) where the CAPs have vanished or in the asymptote
where $\hat{V}_{CAP} \ne 0$.
 {\it As we
show the  optimal CAPs can be introduced also in the region where
the physical potential  has not  vanished}. The un-avoided
reflections due to the use of a finite number of grid points (or
basis functions) are discussed. A simple way to reduce the
"edge-grid" reflection effect is described.
\end{abstract}

\maketitle
\section{Introduction}

There is an extensive use in wave-packet (WP) propagation
calculations in complex absorbing potentials (CAPs). The use of
CAPs in propagation of WP calculations is usually for half
collision experiments where the initial wave-packet is localized
in the interaction region where the CAPs are vanished.  The role
of the CAPs is to avoid the reflection from the edge of the grid
as obtained in the numerical propagation calculations.  Often CAPs
are referred to as optical potentials. The CAPs are used in very
different fields of physics, chemistry and technology. See for
example: calculations of resonances for CAPs in a nuclear physics
problem\cite{Ho}; deriving  new expressions that simplify the
numerical calculations of state-to-state transitions probabilities
for reactive scattering collisions (for time independent
Hamiltonians see Ref.\cite{Miller} and for time dependent ones see
Ref.\cite{Ilya-NM}) ;  calculations of complex molecular potential
energy surfaces by CAPs\cite{Robin}; and molecular electronic
studies where the CAP serves to absorb charge reaching the
electrodes\cite{Roi-Tamar-Danny}. Beside the use of CAPs in the
numerical calculations an effort has been taken in  developing
different type of CAPs. See for example
Ref.\cite{NewCAPS1,NewCAPS2,NewCAPS3,NewCAPS4,NewCAPS5,NewCAPS6,NewCAPS7}
where recently new type of CAPs were developed.  For a most recent
review on CAPs see Ref.\cite{Muga}.

In 1998 we have derived  CAPs  by applying the
smooth-exterior-scaling transformations (SES) to the
Hamiltonian\cite{NM-FRCAP}.   Here we study the  {\it conjecture}
that the use of exterior-scaling or SES similarity
transformations, produce reflection free CAPs (RF-CAPs) for the WP
propagation calculations. As we will show here within the finite
basis-set or finite grid approximations the CAPs are not
reflection free ones. However, it is possible to show that for a
given finite basis/grid method a quantity criteria for the
strength of the numerical reflections can be derived. It is
important to mention that about the same time
 Riss and Mayer\cite{riss} obtained  CAPs,
  which under specific conditions are similar to the SES-CAPs,
  by taking
  another approach (so called Transformative-CAP).  Only when the
CAP is introduced in a region where the  potential energy has been
vanished, the Transformative-CAP derived by Riss and Meyer
\cite{riss} is equal to the SES-CAP that has been derived
analytically without any approximations by us\cite{NM-FRCAP} (in
such a case the SES-CAP and the Transformative-CAP are identical
although they look slightly different). Our main motivation for
deriving SES-CAPs was to simplify the calculations of resonances
positions and widths. However, this SES-CAP has been used also to
avoid the artificial reflections from the edge of the grid in
wave-packet (WP) propagation calculations\cite{WP-FRCAP}. One may
wonder, what is the need for the Transformative-CAPs or the
SES-CAPs  since the reflections can be taken as small as one
wishes by introducing the CAP in the domain where the physical
potential is zero and by making the CAP (any CAP) soft and long
enough\cite{Dieter-Personal}. The answer to that question is that
it is most desired to avoid the use of long ranged CAPs which
require large number of basis functions or large number of grid
points in heavy duty numerical calculations. For example, in
propagation calculations of many electron molecular systems it is
hard to avoid the introduction of the CAP in the domain where the
physical long range potential is not zero.

Here we want to discuss two type of questions. The first type are
mathematical-physical questions (i.e., theoretical questions in
the sense that we assume that complete basis sets are used). Such
as,

- What are the properties of reflection free CAPs ? (as we show
here, there are two conditions that should be satisfied).

- Can we introduce the  RF-CAP in the domain where the physical
potential is not zero and the propagated wave packet does not
consists of out going waves only? (as we show the answer is yes).

- Can the initial state be exponentially localized in the
interaction region,
 as required in half-collision
experiments, where the CAP vanishes? (the answer is yes).

- Can the initial state be localized in the domain where $V_{CAP}
\ne 0$ ? (the answer is yes, provided the smooth-exterior scaling
transformation is applied to the initial state).

 The second type are practical questions.

- Are indeed the RF-CAPs reflection free in the numerical
calculations where
 finite number of grid-points or finite of basis functions are
 used ? (the answer is no since in spite of the complete absorbing of the
fast moving components of the wavepacket still
 there is an edge-grid reflection
 effect which is associated with the slow moving components of the
 WP).

- Can we minimize the reflections which result from the use of
finite sized basis/grid methods and how ? (the answer is yes, by
methods explained in the paper).

- Can we apply the RF-CAPs to many electron problems ? (the answer
is yes provided the electronic repulsion terms,
$1/|\vec{r}_i-\vec{r}_j|$ are modified. This requirement can be
avoided when the ionized electrons are not correlated).

\section{What are the ideal reflection-free CAPs ?}

 First, we
should describe the numerical  problem we want to solve by
introducing a CAP into the Hamiltonian. Using the hermitian
quantum mechanics the propagated wave packet is given by,
\begin{equation}
\Phi_{\it exact}(t)=e^{-{i{\hat H}t}{/\hbar}}\Phi_0.
 \label{exactWP}
\end{equation}

In the numerical calculations the propagated wave packet is
$\Phi_{\it num}(t)\ne \Phi_{\it exact}(t)$.  We are looking for
numerical methods for which,
\begin{equation}
 |\Phi_{\it num}(t)-\Phi_{\it exact}(t)|<\epsilon,
\label{numCONDITION}
\end{equation}
where $\epsilon$ determines the requirement accuracy from the
numerical results. Since in the numerical calculations only finite
number of grid points or finite number of basis functions are
used, the available spatial space is not from $r=0$ to $r=\infty$
but up to  $r=L$. Therefore, accurate results are obtained as long
as $\Phi_{\it exact}(t)$ vanishes at $r\geq L$. By increasing the
number of the grid-points or by increasing the number of the basis
functions we increase the value of L.  The initial state,
$\Phi_0$, is a square integrable function. In half collision
experiments (such in photo-dissociative or auto-ionization
reactions) the initial WP is localized in the interaction region
where $|\Phi_{\it num}(t=0)-\Phi_{\it exact}(t=0)|<\epsilon$.
However, as time passes the wave-packet spreads and only during a
given period of time $\tau$, the numerical calculations satisfy
the accuracy condition stated above. It is important to realize
that the value of $\tau$ is determined by the time it takes for
the tail of the wave-packet to reach the edge of spatial space
(i.e., r=L). In order to obtain $\Phi_{\it num}(t)$ within the
desired accuracy, one should increase the number of the used
grid/basis points/functions and thereby increase the value of L.
{\it The role of the CAP is to enable one to obtain accurate
numerical results in the limited available spatial space, $r \leq
r_{CAP}<L$, without the need to increase the number of grid/basis
points/functions}. Namely,
\begin{equation}
\Phi_{CAP}(t)=e^{-{i({\hat H}+ \hat{V}_{CAP})t}{/\hbar}}\Phi_0,
 \label{CAP-WP}
\end{equation}
where due to the use of the finite grid/basis-set numerical
methods,
\begin{equation}
\Phi_{CAP}(r\geq L,t)=0. \label{PhiCAPeq0}
\end{equation}
The CAP is defined such that,
\begin{equation}
\hat{V}_{CAP}=0 \,\ as \,\ r \leq r_{CAP}<L
 \label{VCAPeq0}
\end{equation}
and,
\begin{equation}
 |\Phi_{CAP}(t)-\Phi_{\it
 exact}(t)|<\epsilon \,\ in \,\ the \,\ region \,\ where \,\ \hat{V}_{CAP}=0.
 \label{PhiCAPeqEXACT}
\end{equation}
A common requirement is that,
\begin{equation}
\Phi_0=0 \,\ in \,\ the \,\ region \,\ where \,\
{\hat{V}}_{CAP}\neq 0
 \label{initialEq0Z}
\end{equation}
As a matter of fact the last condition is too strong and it is
possible to satisfy Eq.\ref{PhiCAPeqEXACT} also when the initial
state is localized in the region where the CAP gets non zero
values. This extension will be discussed later.

Short range CAP
(the Saxson-Wood potential) has been used about two decades ago in
molecular wave-packet calculations\cite{Claude}.  A CAP which has
been used often in the literature
\cite{jolicard1,jolicard2,jolicard3,jolicard4,danny,baer,kouri,neuhauser-kouri,last,jolicard5,uri,volodya,volodya2,neuhauser,diter,lenz}
is $V_0=0$ for $x<x_0$ and $V_0=-i\lambda(x-x_0)^n$ where
$n=1,2,...8$ for $x\geq 0$. For large values of n these CAPs are
very similar to the purely imaginary step type potential that has
been shown above to provide a strong reflection. Regarding the
reflections due to the introducing
 of abrupt complex potentials one might be aware to the fact that there
are examples (see the review in Ref.\cite{Muga}  and references
therein) of discontinuous potentials  that are constructed to
avoid reflection, and absorb totally,  at single incident
energies, or in certain momentum intervals, or at a discrete set
of energies. Of course they cause reflections at other energies.
The CAPs that we are looking for are different ones. They are
energy independent RF-CAPs, and in principle can be chosen to be
universal ones (i.e., problem independent).

As we will show here it is unlikely to have a universal (i.e.,
problem independent) CAP for which both Eq.\ref{PhiCAPeq0} and
Eq.\ref{PhiCAPeqEXACT} are satisfied. Therefore, let us first
discuss the possibility to satisfy Eq.\ref{PhiCAPeqEXACT} when the
condition given by Eq.\ref{PhiCAPeq0}  is replaced by a weaker
numerical condition: $\Phi_{CAP}(t)$ is a square integrable
function at any given time, which decays to zero much {\it faster}
than the exact solution. Such that within a given time interval,
\begin{equation}
 \Phi_{CAP}(r=L(T),t<T)\leq \epsilon,
 \label{PhiCAPsquareintegrable}
\end{equation}
where the value of $\epsilon$ is determined from  the desired
accuracy of the numerical calculations.

\section{The SES-transformations and the conditions for optimal
reflection-free CAPs.}


 The idea of introducing
RF-CAPs by using the exterior scaling or SES methods is clear: the
Hamiltonian remains as it is inside the inner region, where the
coordinates stay on the real axis. However, it has been shown by
Simon that upon the exterior scaling transformation,
\begin{equation}
r\to r_{ext}
\end{equation}
where inside the inner unscaled region,
\begin{equation}
r_{ext}=r \,\ when \,\ r \leq r_{CAP}
\end{equation}
and in the external-scaled region,
\begin{equation}
r_{ext}=r_{CAP}+(r-r_{CAP})e^{i\theta} \,\ when \,\ r>r_{CAP}
\end{equation}
 the  eigenfunctions are {\it not equal} to eigenfunctions of the unscaled (i.e.,
hermitian) problem {\it inside} the unscaled region\cite{Simon}.
For example, for a free particle Hamiltonian the continuum
eigenfunctions inside the inner unscaled region are given by,
$A_{in}\exp(-ik\exp(-i\theta)r)+A_{out}\exp(+ik\exp(-i\theta)r)$.
Since the propagated WP can be described as a linear combination
of the eigenfunctions of the complex scaled (or exterior scaled)
Hamiltonian, it is not clear at all that in this case
Eq.\ref{PhiCAPeqEXACT} is satisfied (here we consider the exterior
scaled Hamiltonian as $\hat{H}+\hat{V}_{CAP}$). This result is
very confusing since from numerical propagation calculations we
know that inside the inner unscaled region in space, the
propagated WP is exactly as obtained without the use of
exterior-scaling. As we will show below the validity
Eq.\ref{PhiCAPeqEXACT} can be easily explained by association the
SES approach with the use of similarity transformation operators
as developed in Ref.\cite{NM-FRCAP,NM-review}. Using the SES
approach,
\begin{equation}
r\to r_{SES}\equiv F_{\theta}(r),
\end{equation}
where the path in the complex coordinate space is chosen such
that,
\begin{equation}
|F_{\theta}(r)-r|\leq \epsilon \,\ when \,\  r < r_{CAP}
\label{SES-inner}
\end{equation}
and,
\begin{equation}
 \frac{F_{\theta}(r)}{r} \rightarrow e^{i\theta} \quad as \quad r
\rightarrow \infty.
 \label{F-function}
\end{equation}
The SES transformations clearly show that Eq.\ref{PhiCAPeqEXACT}
can be satisfied to any desired accuracy. If the SES
transformation is represented by the similarity operator, ${\hat
{S}}$, than the propagated WP within the framework of the SES
approach is given by ${\hat S}\Psi_{exact}(t)$ which is equal to
$\Psi_{exact}(t)$ inside the inner region (see Eq.\ref{SES-inner})
where ${\hat S} \sim 1$.

 Let us discuss now the validity of Eq.\ref{PhiCAPsquareintegrable}.
Following Simon's proof for the exterior scaled potential and
following Moiseyev and Hirschfelder's proof for general complex
scaled transformations\cite{NM-JOH} (including the SES
transformations), the complex scaled resonances functions are
square integrable but the continuum eigenfunctions are not square
integrable functions. They are associated with complex
eigenvalues,
$E_{ext}(continuum)=k_{ext}^2/2=(k\exp(-i\theta)^2/2$. Such that
$k_{ext}\cdot r_{ext}$ in the {\it exterior} region is equal to
the same value as obtained in hermitian quantum mechanics, i.e.,
$k\cdot r$ (note that a very different result is obtained in the
inner region as discussed in the previous paragraph).  Therefore,
the asymptote of the continuum wavefunctions as obtained after the
application of the exterior or SES transformations, remain as
obtained within the framework of the conventional (i.e.,
hermitian) QM approach. Upon complex scaling $k_{ext}$ is rotated
into the lower-half complex k-plane to avoid the exponentially
divergence of the complex scaled incoming waves associated with
real and positive values for the wave vector, i.e.,
$\exp(-ikr_{ext})=
\exp(-ik\cos(\theta)r)\exp(+k\sin(\theta)r)\to\infty$, as $r\to
\infty$. Therefore, it is no obvious weather a square integrable
WP such as, $\Phi_{exact}(t)=\int_{-\infty}^{+\infty}
C(k,t)\exp(ikr) dk \to 0$ as $r\to\infty$, remains square
integrable when $r\to r_{ext}$ or $r\to r_{SES}$. It has been
proven by Moiseyev and Katriel\cite{NM-Katriel} that for
sufficiently small values of $\theta$, i.e., $\theta<\theta_c$,
the eigenfunctions of a complex scaled  Hamiltonian which are
associated with the bound states are square integrable. The value
of $\theta_c$ depends on the shape of the
potential\cite{NM-Katriel}. Let us assume that the wave-packet is
a Gaussian, $\exp(-ar^2)$. It is clear that
$\exp(-ar_{ext,SES}^2)$ remains square integrable provided that,
$\theta\leq \theta_c=\pi/2$. When the wave-packet is more
localized, for example is described as $\exp(-ar^N)$, then
$\theta_c=\pi/N$. Since Gaussians form an over-complete basis set,
one might expect that any square integrable function (which can be
expanded in term of the Gaussian basis set) remains square
integrable after applying the complex scaling or the SES
transformation.

 It
is easy to prove that the wave-packet $\Phi_{exact}(t)$ decays
exponentially to zero at any given time, provided it is a square
integrable function at t=0. A proof which holds also for complex
scaled non-hermitian Hamiltonians is as follow:
$\Phi_{exact}(t+dt)= \exp(-i\hat{H}dt/\hbar)\Phi_{exact}(t)$. For
sufficiently small value of dt,
$\exp(-i\hat{H}dt/\hbar)=\sum_{n=0}
(n!)^{-1}(-idt/\hbar)^{n}(\hat{H})^{n}$ is a converged series
(provided $Im {\hat H}\le 0$). If $\Phi_{exact}(t'\leq t)$ is a
square integrable function than $(\hat H)^{n}\Phi_{exact}(t)$ is
square integrable as well (the second derivative of a square
integrable function is square integrable and the product of a
square integrable function and a confined (complex scaled)
potential is also a square integrable function).

Let us summarize the facts we know by now: (1) when the initial
wave-packet (WP) is square integrable the time propagated WP is
square integrable as well; (2) the complex scaled square
integrable WP remains square integrable; (3) the complex scaled
incoming waves diverge exponentially whereas the outgoing waves
exponentially decay  to zero; (4) in the absence of a source of
particles in infinite large distance from the studied system, the
asymptote of the propagated WP consists of out going waves only
(as in half collision experiments). From (1)-(3) it is clear that
for the most general case the square integrable WP,
\begin{equation}
 \Phi_{exact}(r\geq L,t) = \int_0^\infty dk (D(k,t)e^{-ikr}+C(k,t)
 e^{+ikr}) \to 0\,\ as \,\ r\to \infty,
\end{equation}
remains square integrable,
\begin{eqnarray}
 \Phi_{CAP}(r\geq L,t) &=& \int_0^\infty dk (D(k,t)e^{-ik\cos(\theta)r}e^{+k\sin(\theta)r}+ C(k,t)
 e^{+ik\cos(\theta)r}e^{-k\sin(\theta)r}) \\
 &\to& 0\,\ as \,\ r\to \infty,
\end{eqnarray}
although each one of the components of the complex scaled incoming
waves exponentially diverge. When the condition (4) is not
satisfied this fact (i.e., interference of exponentially diverged
incoming waves results in a square integrable function) may
introduce some numerical difficulties in the propagation
calculations. For overcoming these type of numerical difficulties
when long ranged potentials are used see the second reference in
Ref.\cite{uri}.

When condition (4) is satisfied (as in all half collisions
experiments) then,
\begin{equation}
 \Phi_{exact}(r\geq L,t)=\int_0^\infty dk C(k,t) e^{+ikr}\to
 0\,\ as \,\ r\to \infty,
 \label{asymptotePhi}
\end{equation}
and it is easy to see that $\Phi_{CAP}$ decays faster since,
\begin{equation}
 \Phi_{CAP}(r\geq L,t)=\int_0^\infty dk C(k,t)
 e^{+ik\cos(\theta)r}e^{-k\sin(\theta)r}.
 \label{asymptotePhiCAP}
\end{equation}
The fact that  within the interval of $r_{CAP}<r\leq L$, the
propagated WP, $\Phi_{CAP}(t)$, decays {\it faster} than
$\Phi_{excat}(t)$, is the main motivation behind the use of the
exterior scaling, smooth-exterior complex scaling methods in the
numerical propagation calculations.

\section{ A quantity  criteria for the measurement of the strength of
the numerical reflections}


 From Eq.\ref{asymptotePhiCAP} a quantity criteria
for the strength of the numerical reflections from the edge of the
grid is obtained,
\begin{equation}
 \left |C(k,t)e^{-k\sin(\theta)L}\right | \leq \epsilon.
 \label{Criterion}
 \end{equation}
As an upper limit for the accuracy of the calculations one gets
that,
\begin{equation}
 \left |C(k,t)e^{-kL}\right | \leq \epsilon.
 \label{UPPER-LIMIT}
\end{equation}

At t=0 the initial wave-packet gets exponentially small values at
$r\geq L$ and therefore we can consider it as a case where,
$C(k,0)=0$. As time passes the wavelet with the largest value of k
(associated with a large velocity) is the first to reach the edge
of the grid. As one can see from Eq.\ref{UPPER-LIMIT} the fast
moving components of the wavepacket are entirely absorbed at
$r=L$, due the use of the complex absorbing boundary conditions
which were introduced by the use of the exterior scaling or the
SES transformations. For the components of the wavepacket
associated with small values of k, the requirement of
$\exp(-kL)\sim 0$ is satisfied by increasing the value of L. The
propagation calculations using SES transformations, within the
framework of the finite basis-set/grid approximations, are
accurate as long as $|C(k,t)|$ gets sufficiently small values.
This explains why L in Eq.\ref{PhiCAPsquareintegrable} is a
function of time and why the duration of the propagation
calculations can not exceed a given period of time, T, when L is
held fixed in the propagation calculations.

As an illustrative example we carried out wavepacket propagation
calculations for a one-dimensional Gaussian, $\Psi(x,t=0)=
(1/5\pi)^{1/4}\exp(-x^2/10 + ip_0x)$, which is localized at a
potential well embedded in between two identical potential
barriers. This potential, $V(x)=(0.5x^2-0.8)\exp(-0.1x^2)$, has
been used before as a test problem for new methods developed in
non-hermitian quantum mechanics (see for example \cite{NM-review}
and references therein).
\begin{figure}
\includegraphics[width=3.in]{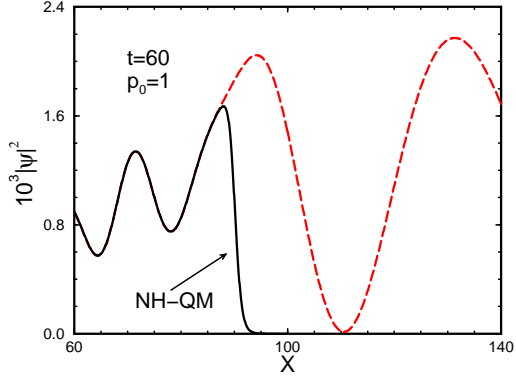}
\caption{\label{Fig1} The numerical exact propagated wavepacket
(long dashed line), $\Psi_{exact}(x,t=60)$, and the corresponding
wavepacket (denoted by NH-QM) which is defined as
$\Psi(F_{\theta=0.5\,rad}(x),t=60)$. The smooth-exterior-scaling
contour is defined as, $F_{\theta} \sim x$ when $|x|<x_{CAP}\equiv
90$, whereas $F_{\theta}=x\exp(i\theta)$ when $|x|>x_{CAP}$. The
initial wavepacket is given by, $\Psi(x,t=0)=
(1/5\pi)^{1/4}\exp(-x^2/10 + ip_0x)$, where $p_0=1$.}
\end{figure}
 In Fig.~1 the results obtained from two
type of propagation calculations are presented. The long dashed
line stands for the numerically exact calculations of
$\Psi_{exact}(x,t=60)$, using 5-order split operator with
$-1000\leq x \leq +1000$. The full solid line is
$\Psi(F_{\theta=0.5\, rad}(x),t=60)$ where $F_{\theta}(x)$ is a
smooth exterior scaling function. Such that, $F_{\theta} \sim x$
when $|x|<x_{CAP}\equiv 90$, whereas $F_{\theta}=x\exp(i\theta)$
when $|x|>x_{CAP}$.  For $|x|<x_{CAP}$
$\Psi_{exact}(x,t=60)\approx \Psi(F_{\theta}(x),t=60)$. However,
it is clearly shown that unlike the exact wavepacket which
oscillates, the smooth exterior scaled wavepacket (labelled in
Fig.1 by non-hermitian quantum mechanics (NH-QM) approach) decays
to zero as  x is rotated into the complex coordinate plane around
$x=x_{CAP}=90$.
\begin{figure}
\includegraphics[width=3.in]{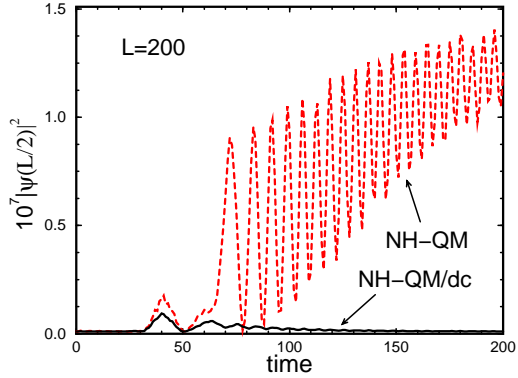}
 \caption{\label{Fig2} {The non-hermitian propagated
wavepacket which is constructed from 400 Fourier basis functions
(long dashed line, denoted by NH-QM) as function of time at
$x=L/2$ (edge of the grid). The propagated wavepacket is defined
as $\Psi(F_{\theta=0.5\,rad}(x=100),t)$. Such that, $F_{\theta}
\sim x$ when $|x|<x_{CAP}\equiv 90$, whereas
$F_{\theta}=x\exp(i\theta)$ when $|x|>x_{CAP}$. The initial
wavepacket is given by, $\Psi(x,t=0)= (1/5\pi)^{1/4}\exp(-x^2/10 +
ip_0x)$, where $p_0=0$. The reflections from the edge of the grid
as time passes are obtained when $t>T\equiv 30$. The full line
stands for the results obtained when a dc-field has been
introduced close to the edge of the grid, $x_{dc}=95$ and ${\cal
E}_{dc}=2$.}}
\end{figure}

Following our analysis the propagated wavepacket decays to zero
when the contour x is smooth exterior scaled (rotated) into the
complex coordinate space only within the time interval $t\leq T$.
The results presented (denoted by NH-QM) in Fig.2 were obtained
from numerical calculations where $-100\leq x\leq+100$ (i.e., the
box size is $L=200$). It is clearly shown that until $t\leq 30$
the complex scaled wavepacket is practically equal to zero at the
edge of the grid (i.e., at x=L/2). As time exceeds the value of
$t=T\equiv 30$ the complex scaled wavepacket is reflected from the
edge of the grid.

\section{How to reduce the numerical reflections of the slow moving
components of the wavepacket from the edge of the grid ?}

Let us propose two different possibilities:

 {\it (a) Accelerate the slow moving components of the wavepacket by inducing an external
dc-field.}

Eq.\ref{UPPER-LIMIT} indicates clearly that the numerical
reflections  from the edge of the grid, are associated with slow
moving wavelengths.  As discussed above the fast moving components
of the wavepacket are entirely absorbed at $r=L$, due the use of
the SES-CAPs. A possible solution to this problem is by adding a
static field close to the edge of the grid, in order to accelerate
the slow moving wavelengths which are completely absorbed by the
SES-CAPs. The static field is turned on only at the edge of the
grid where $r\geq r_{dc}$ and is given by,
\begin{equation}
V_{ext-dc}(r\geq r_{dc})=-\frac{{\cal E}_{dc}}{2}(r-r_{dc}).
\end{equation}

The value of $r_{dc}$ and ${\cal E}_{dc}$ can be optimized to
minimize the effect of the dc-field on the WP propagation. An
estimate of the error introduced by adding the $V_{ext-dc}$
potential term to the Hamiltonian can be obtained from the
imaginary parts of the bound states calculated for the Hamiltonian
which is taken as, ${\hat H} + V_{ext-dc} + {\hat V}_{CAP}$. The
evaluation of ${\hat V}_{CAP}$, from the SES transformations will
be described in Section 6.

Let us return to our illustrative numerical example. The smooth
exterior scaled wavepacket has been calculated as described above
when $|x|\leq 100$ (i.e., the box size is $L=200$ ) when
$V_{ext-dc}$ dc-potential term was added into the Hamiltonian. The
results presented in Fig.~2 (the full solid line denoted by
NH-QM/dc) clearly show the strong suppression of the reflections
from the edge of the grid as the dc field has been introduced into
the Hamiltonian.
\begin{figure}
\includegraphics[width=3.in]{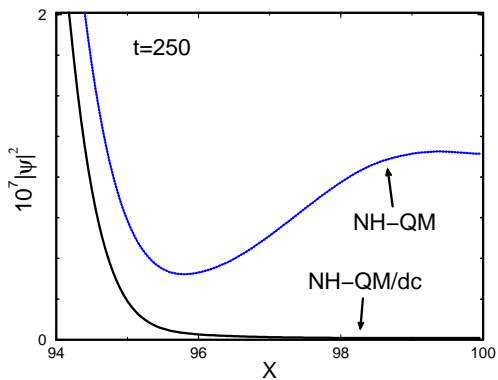}
\caption{\label{Fig3} {The numerical propagated wavepackets at
t=250 as obtained when 400 Fourier functions were used as a basis
set. The reflections from the edge of the grid (which appear at
$t>30$ as shown in Fig.~2) are avoided as a dc field is added
close to the edge of the grid.}}
\end{figure}
In Fig.~3 we present the numerical results
obtained at $t=250$. It is clearly shown that close to the edge of
the grid the dc field inhibits  the artificial reflections as
appeared in the NH-QM calculations. Note that as time passes the
reflection leads to the distortion of the wavepacket also at
regions which are quite far from the edge of the grid, as it is
shown in Fig.~4. However, as one can see from the results
presented in Fig.~4 the introducing of the static field reduces
this artificial edge-grid reflection effect.

\begin{figure}
\includegraphics[width=3.in]{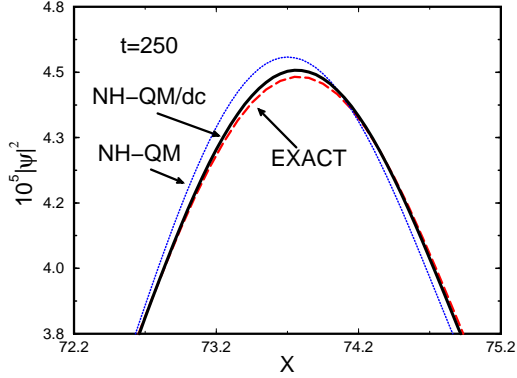}
\caption{\label{Fig4} {The propagated wavepackets at t=250 as
obtained when 400 Fourier functions were used as a basis set, in a
comparison with the numerically exact solution.   The propagated
wavepacket as obtained when both the smooth exterior scaling
transformation and the dc field were implemented into the
numerical calculations, is in a very good agreement with the
numerical exact solution.}}
\end{figure}

{\it (b) Imposing of out-going boundary conditions}

 The numerical
edge-grid reflection effect can be reduced by imposing out-going
boundary conditions (complex scaled ones in our case). It is
simple to implement that approach when grid methods are used,
\begin{equation}
 {\vec \Phi}_{CAP}(t+\Delta t) = {\bf U}
{\vec \Phi}_{CAP}(t),
\end{equation}
where ${\bf U}(t+\Delta t \leftarrow t)$ is the time evolution NxN
matrix and the jth component of the vector ${\vec \Phi}_{CAP}(t)$
is the value of the propagated WP at ${\vec r}_j; j=1,2,..N$ grid
point. The grids points are ordered such that $|\vec{r}_1|\le
|\vec{r}_2|\le ... |\vec{r}_{N-1}|\le |\vec{r}_N|$. The notation
of CAP stands for the use of the exterior or the SES
transformations. The out-going boundary conditions are imposed by
replacing the N-th row of the time evolution matrix $U_{N,i};\,
i=1,2,...,N$ by
$\exp[i\exp(i\theta)\vec{k}(\vec{r}_N-\vec{r}_{N-1})]U_{N-1,i}$.
Here we use the fact that the ${\vec r}_N,{\vec r}_{N-1},{\vec
r}_{N-2}$ grid points are in the scaled region where
$\vec{r}\rightarrow \vec{r}\exp(i\theta)$. The wave vector ${\vec
k}$ is determined from the previous time step calculations and
from the $\vec {r}_{N-1}$ and the $\vec{r}_{N-2}$ grid points.
That is,
\begin{equation}
\exp[{i{\exp(i\theta)\vec{k}(\vec{r}_{N-1}-\vec{r}_{N-2})}}]
=\frac{\Phi_{CAP}(\vec{r}_{N-1},t)}{\Phi_{CAP}(\vec{r}_{N-2},t)}.
\end{equation}
It should be stressed that it is not always true that at a given
time the tail of the WP is constructed of a single out-going wave
component. However, this kind of an approximation has been found
useful in WP propagation calculations of various physical
problems\cite{IEEE}.

 One should assure that the real part of the wave
vectors get positive values only.
 In the
one-dimensional case where equally spaced grid points are used,
the application of that approach is straightforward. In such a
case the  modified last row of the time evolution matrix is given
by,
\begin{equation}
U_{N,i}(t+\Delta t \leftarrow t) =U_{N-1,i}(t+\Delta t \leftarrow
t)\frac{\Phi_{CAP}({r}_{N-1},t)}{\Phi_{CAP}({r}_{N-2},t)}.
\label{OG}
\end{equation}
Similarly, one can modify all $j>j_c$ rows and not only the last
one. The assumption is that the vectors, ${\vec {r}_{j>j_c}}$ are
all embedded in the asymptote region of the propagated wavepacket.
 This method
(applicable to 3D problems as well) to reduce the edge-grid
reflection effect is an extension/variation of Hadley's original
work, where the transparent boundary condition for beam
propagation method was developed\cite{IEEE}. This  method does not
require the use of CAPs. However, we believe that the use of the
exterior or the SES transformation together with the transparent
boundary condition, should minimize the numerical reflections from
the edge of the grid.

\begin{figure}
\includegraphics[width=3.in]{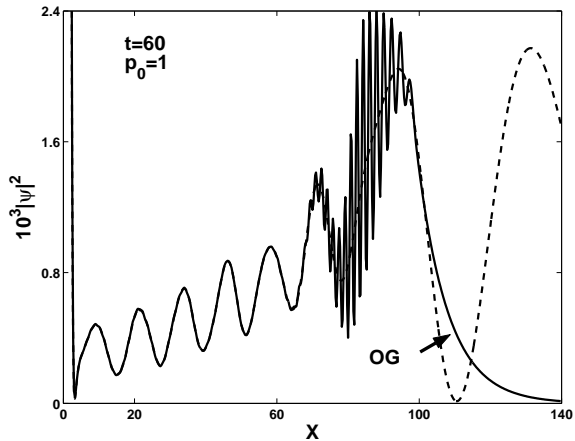}
\caption{\label{Fig5} {The propagated wavepacket as obtained when
the time evolution operator has been modified as explained in the
text (see Eq.\ref{OG}). The model Hamiltonian  and the initial
state are as described in the caption of Fig.~1. The long dashed
line stands for the results obtained from numerically exact
propagation calculations. }}
\end{figure}
The possibly to impose outgoing boundary condition by modifying
the time evolution operator as shown in Eq.\ref{OG}, is
illustrated here by applying it to our test-case  problem. The
results presented in Fig.~5 clearly show that by using the method
introduced above a similar absorbing boundary condition effect -
as achieved when the RF-CAPS are added to the Hamiltonian - is
obtained. A comparison between  the results presented in Fig.~5
and Fig.~1 shows that the RF-CAP (Fig.~1) provides better results
than those obtained by imposing out going boundary condition on
the propagated wavepacket (Fig.~5). However, it might be expected
that similarly to the effect of the dc-field described above, the
combination of the two approaches would avoid the reflections
which are obtained after long time propagation. This study is out
of the scope of the present study and requires a further
investigation.

\section{The optimal RF-CAPs from the SES transformations.}

In order to complete the representation of optimal RF-CAPs we
should show that under well defined specific conditions the use of
the SES transformation, is equivalent to the inclusion of a CAP
which gets non-zero values only in the edge of the grid. This
SES-CAP is an non-local operator, since it includes terms with the
momentum and kinetic energy operators. For the sake of a coherent
representation of the subject we briefly describe how the SES-CAPs
are obtained.

The SES transformed time dependent Schr\"odinger equation  can be
rewritten as,
\begin{equation}
{\cal H}_{CAP} \Phi_{CAP}(t) = i\frac{\partial}{\partial t}
\Phi_{CAP}(t) \label{NH-TDSE}
\end{equation}
where,
\begin{equation}
 {\cal H}_{CAP}={\hat S}{\hat H}{\hat S}^{-1}
 \label{S-HAM}\\
\end{equation}
\begin{equation}
 \Phi_{CAP} = {\hat S}\Phi_{\it exact}(t)=\Phi_{\it
 exact}(F_{\theta}(r),t)
\label{phiCAP}\\
\end{equation}

 We have proved before that the SES transformation is equivalent to the including of a non-local
energy independent, universal (i.e., problem independent)
CAP\cite{NM-FRCAP},
\begin{equation}
{\cal H}_{CAP}= {\hat H} + \Delta V + {\hat V}_{RF-CAP},
 \label{HAM-CAP}
\end{equation}
where the correction term to the physical potential is given by,
\begin{equation}
\Delta V = V(F_{\theta}(r))-V(r),
 \label{DV-CAP}
\end{equation}
and the  non-local energy independent, universal  CAP has been
proved to be equal to\cite{NM-FRCAP},
\begin{equation}
{\hat
V}_{RF-CAP}=V_0(r,\theta)+V_1(r,\theta)\frac{\partial}{\partial
r}+V_2(r,\theta)\frac{\partial^2}{\partial r^2}
 \label{POT-CAP}
\end{equation}
The complex functions $V_j\,; j=0,1,2$ are vanished in the inner
region where $r<r_{CAP}$. They are inverse proportional to the
mass (reduced mass) of the particle (s) which is (are) absorbed by
the ${\hat {V}}_{RF-CAP}$ and are defined as ( note that below we
use the notation $F^{(n)}_\theta(r)\equiv dF^n_{\theta}(r)/dr^n$)
\cite{NM-FRCAP}:
\begin{equation}
V_0(r,\theta)=\frac{\hbar^2}{4M(F^{(1)}_\theta(r))^3}\left[F^{(3)}_\theta(r)-\frac{5(F^{(2)}_\theta(r))^2}
{2F^{(1)}_\theta(r)} \right]
 \label{V0-CAP}
\end{equation}
\begin{equation}
V_1(r,\theta)=\frac{\hbar^2F^{(2)}_\theta(r)}{M(F^{(1)}_\theta(r))^3}
 \label{V1-CAP}
\end{equation}
\begin{equation}
V_2(r,\theta)=\frac{\hbar^2}{2M}\left(1-(F^{(1)}_\theta(r))^{-2}\right)
\label{V2-CAP}
\end{equation}
The initial state is defined as,
\begin{equation}
\Phi_0(transformed)= \Phi(F_{\theta}(r),t=0)
 \label{initial}
\end{equation}

When the initial state is localized in the interaction region
where ${\hat {V}}_{RF-CAP}=0$ then
$\Phi(F_{\theta}(r),t=0)=\Phi(r,t=0)$. In the case that the
physical potential is a short range potential it is quite obvious
that the contour of integration $F_{\theta}(r)$ can be chosen to
yield $\Delta V =0 $ everywhere at any point in the entire space.
In the case of long range potential the situation is more
complicated\cite{Shachar}. In such a case the $\Delta V$ term in
Eq.\ref{HAM-CAP} can not be neglected and the SES RF-CAP is equal
to $ \Delta V + {\hat V}_{RF-CAP}$ and seems to be problem
dependent. {\it However, for neutral molecules if $r_{CAP}$ gets a
sufficient large value such that the ionized electrons are in
hydrogenic like orbitals  then $ \Delta V + {\hat V}_{CAP}$ can be
replaced by a universal potential term} $1/r-1/F_{\theta}(r)
+{\hat V}_{RF-CAP}$\cite{Shachar}. This approach holds also for
many electron systems where we assume that the ionized electrons
are not correlated as they get far away from the atom/molecule/QD.
In such a case  we do not need to replace the two electron
repulsion terms $|\vec{r}_i-\vec{r}_j|^{-1}$ by
$|F_{\theta}(\vec{r}_i)-F_{\theta}(\vec{r}_j)|^{-1}$.

Before concluding let us return to our illustrative numerical
example. Using 400 Fourier  basis functions  (with the box-size,
L=200) we obtained matrix representations of the hermitian
Hamiltonian, ${\hat
H_{H-QM}}=-0.5d^2/dx^2+(0.5x^2-0.8)\exp(-0.1x^2)$, and also of the
non-hermitian one, ${\hat H}_{NH-QM}={\hat H_{H-QM}}+{\hat
V}_{RF-CAP}$. The parameters for the function $F_{\theta}(x)$ as
defined in Ref.\cite{NM-FRCAP,WP-FRCAP} are $\theta=0.5 rad$,
$\lambda=0.9$ and $x_{CAP}=90$. The two 400x400 matrices where
diagonalized. The eigenvalues and eigenvectors of the Hermitian
matrix are correspondingly given by, $E_j^{H-QM}(real)$ and
${\vec{C}}_j^{H-QM}$. Similarly, $E_j^{NH-QM}(complex)$ and
${\vec{C}}_j^{NH-QM}$ are associated with the eigenvalues and
eigenvectors of the non-hermitian (complex and symmetric) matrix.
The propagated wavepacket within the framework of H-QM is given
by,
\begin{equation}
\Psi_{H-QM}(x,t)=\sum_{j=1}^{400}
e^{-iE_j^{H-QM}t}\sum_nC_{n,j}^{H-QM} \exp(i\frac{2\pi n x}{L})
\end{equation}

The propagated wavepacket within the framework of the NH-QM
approach is given by,
\begin{equation}
\Psi_{NH-QM}(x,t)=\sum_{j=1}^{400}
e^{-iE_j^{NH-QM}t}\sum_nC_{n,j}^{NH-QM} \exp(i\frac{2\pi nx}{L})
\end{equation}

\begin{figure}
\includegraphics[width=3.in]{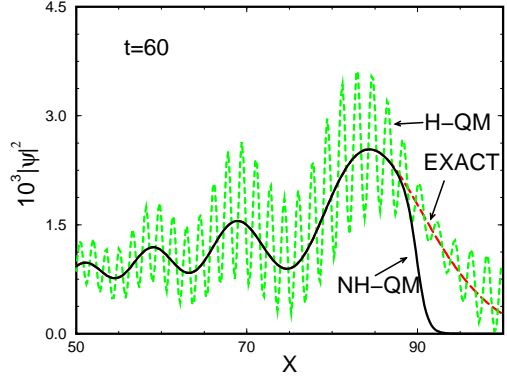}
\caption{\label{Fig6} {The  propagated wavepackets at t=60 as
obtained from conventional and non-hermitian QM calculations, in a
comparison with the numerically exact solution (denoted by a long
dashed line). The propagated wavepackets denoted by H-QM and NH-QM
correspondingly were constructed from the eigenfunctions and
eigenvalues of the hermitian and non-hermitian Hamiltonians when
400 Fourier basis functions were used as a basis set. The NH-QM
results obtained when the smooth-exterior-scaling transformation
was introduced ( by adding ${\hat V}_{RF-CAP}$ into the Hermitian
hamiltonian), are in a complete agreement with the exact solution
when  $|x|<x_{CAP}\equiv 90$.}}
\end{figure}
The results presented in Fig.~6 clearly show that while the
reflections from the edges of the grid appeared in the propagation
calculations within the framework of the conventional QM, they do
not show up in the NH-QM calculations. The reflections appeared in
the conventional quantum mechanical calculations  due to the use
of the eigenfunctions, which were obtained within the framework of
the box-quantization approximation, as a basis set. A
quasi-discrete continuum rather than a continuous continuum  has
been used in the propagation calculations. As one can see from the
results presented in Fig.~6 the use of the RF-CAP provides,
$\Psi_{NH-QM}(-90<x<+90,t=60)=\Psi_{exact}(-90<x<+90,t=60)$
(within more than 6 digits of accuracy).  In spite of the fact
that in the two calculations we have used the same basis functions
and the same number of them the NH-QM calculations provided an
accurate propagated wavepacket while the conventional calculations
are far from convergence.
\section{Concluding remarks}
We can summarize it by saying that for the CAPs derived from the
SES  transformations\cite{NM-FRCAP}: (1) the propagated WP decays
faster to zero than the exact solution and therefore at any given
time we can use a smaller grid/basis in the numerical calculations
when the SES-CAPs are introduced into the numerical calculations;
(2) the SES-CAPs can be introduced also in the region where the
interaction potential is active (provided the edge of the grid is
in the region where the exact WP has outgoing wave components
only); (3) the use of SES-CAPs enables one to introduce the CAPs
also in the region where the initial WP does not get zero values;
(4) the duration of the WP calculations which provide accurate
results (avoiding the numerical reflections from the edge of the
grid) can be easily estimated (see Fig.~3); (5) it is possible to
reduce the reflections of the slow moving components of the
wavepacket, either by introducing a dc-field in the edge of the
grid or by imposing out-going boundary conditions on the
propagated WP. (6) The SES-CAPs are indeed the optimal reflection
free caps, RF-CAPS, for wavepacket propagation calculations.

\begin{acknowledgments}

This work was supported in part by the by the Israel Science
Foundation (grant no. 73/01) and by the Fund of promotion of
research at the Technion. One of us (NM) acknowledge Juan G. Muga
(the correspondence author of a recent review on CAPs) from Bilbao
for drawing his attention to the open question: what are the
perfect reflection free CAPs for motion of wave packet
calculations ? It is a pleasure to thank Hans-Dieter Meyer from
Heidelberg for his most helpful comments and Yoav Berlatzky from
the Technion for drawing our attention to Ref.[39].

\end{acknowledgments}

\end{document}